\documentclass[11pt,a4paper]{article}

\usepackage[T1]{fontenc}
\usepackage[utf8]{inputenc}
\usepackage[sc,osf]{mathpazo}
\usepackage[a4paper,textwidth=138mm,textheight=215mm,top=32mm]{geometry}
\usepackage{microtype}
\usepackage{titlesec}
\usepackage{fancyhdr}
\usepackage[hidelinks]{hyperref}

\titleformat{\section}{\large\scshape}{\thesection.}{0.6em}{}
\titlespacing*{\section}{0pt}{1.6em}{0.5em}

\hypersetup{
  pdftitle={Ignore the Writing: A Proposal for Peer Review in the Age of AI},
  pdfauthor={Vassilis Kostakos},
  pdfkeywords={Peer Review, Generative AI, Research Assessment, Publication Models, Meta-research}
}

\begin{document}
\thispagestyle{empty}

\begin{center}
  {\LARGE\bfseries Ignore the Writing: A Proposal for Peer Review in the Age of AI\par}
  
  \vspace{1.6em}
  {\large Vassilis Kostakos\par}
  \vspace{0.4em}
  {\normalsize School of Computing and Information Systems\\
   The University of Melbourne, Australia\par}
  \vspace{0.3em}
  {\small\href{mailto:vassilis.kostakos@unimelb.edu.au}{vassilis.kostakos@unimelb.edu.au}\par}
  \vspace{0.8em}
  {\small October 2026\par}
\end{center}

\vspace{1.2em}

\begin{center}
  \begin{minipage}{0.9\textwidth}
    \begin{center}\scshape Abstract\end{center}
    \vspace{0.3em}
    \noindent\small
    The written paper has long stood in as proof of work: a fluent, well-structured manuscript was taken as evidence that good research lay behind it. Generative AI has broken that link. Polished manuscripts now cost nothing to produce, submissions are outgrowing what reviewers can absorb, and venues are turning to AI for help. The logical endpoint is a loop where AI writes the papers and AI reviews them. We argue that this loop is survivable, but only if it is given a reward function, and that human reviewing effort should become that reward function. We propose that submissions should consist of a 1500-word ``brief'' plus an unlimited, machine-verifiable appendix. Human reviewers judge whether a pursuit is worthwhile, models verify the experimental record in the appendix, and human curation decides where the field goes next.

    \vspace{0.6em}
    \noindent\textit{Keywords:} peer review, generative AI, research assessment, publication models, meta-research
  \end{minipage}
\end{center}

\vspace{1.4em}

\section{Introduction}
The academic paper has long doubled as proof of work. A polished manuscript signalled months of effort, and reviewers treated clarity of prose as a proxy for clarity of thought, and thoroughness of description as a proxy for thoroughness of method. The conflation was harmless while writing well was expensive. Generative AI has now ended this correlation. Length is now free and competent academic prose is available on demand, so text quality must be assumed uniformly high across submissions. The two properties reviewers have leaned on, how much was written and how well, no longer separate good work from bad.

We argue for something counter-intuitive: writing should be taken out of the review process altogether, and reviewers should, for the most part, ignore it.

\section{Background: Text as Proof of Work}
Our reviewing apparatus was calibrated for a world in which text is the honest residue of intellectual labour. But now a single author can generate more plausible-looking manuscripts in a weekend than a subcommittee can read in a cycle. Responses from journals and conferences so far have been sensible enough, mostly AI tools that help humans decide faster, but the trajectory is clear: we will soon be relying on fully machine-generated reviews. Because AI produces text faster than humans can read it, in the long run the equilibrium is AI writing papers and AI reviewing them. We should plan for that endpoint now.

\section{The Mutual Automation Fallacy}
Research communities have reacted to AI with a peculiar symmetry. Qualitative researchers suspect the quantitative tradition is finished: software and analysis can be automated, freeing everyone for the genuinely interesting questions. Quantitative researchers suspect the reverse: a model will happily turn transcripts into themes, freeing everyone to build useful technical solutions. Everyone is confident that AI will replace everyone else, just not themselves. Both camps are half right. What can be automated is the mechanical middle of each tradition, not its ends: deciding what is worth studying, and deciding whether a solution is actually desirable in the world. This returns us to where most disciplines have always been, with qualitative researchers identifying interesting problems and quantitative researchers solving them. What is disappearing is the artefact that sat between them, the long human-written paper. That is a problem, because the paper is the only thing our review process knows how to assess.

\section{What Should a Venue Review?}
Review has been, above all, a stamp of approval on a work's quality, conduct, and communication. But close reading is no longer a sensible use of our scarce intellectual resources when much of the text is machine-generated. To review the quality of generated prose is to review the machine, and prose whose polish bears no relation to what actually happened cannot certify how the work was conducted. The real question for a venue, then, is not how to review more papers, but what it should be reviewing at all.

The AI-writes-AI-reviews loop has one genuine virtue. If models read, write, extend and re-run the literature, contributions iterate at a pace no human pipeline could match. But any self-improving loop needs a reward function. There is a harder physics constraint too: AI can explore a boundless space of possibilities, but every generated paper and verification run uses our finite energy and compute. Thus, prioritisation becomes a necessity, and the question is whether it happens by accident or by design. Human reviewing should become each discipline's reward function and curation of where contributions are heading.

\section{A Proposal: The 1500-Word Brief}
We propose that papers become a 1500-word brief with an unlimited appendix. The brief presents the idea and makes the case for its value. The appendix carries procedures, results, analyses and data at whatever length they require, along with related work, discussion, and references. Machine learning conferences have run short page limits with unlimited appendices for years, and their reviewing crisis is arguably worse than anyone's, because they changed the length of the text without changing what humans do with it. The substance of our proposal is therefore not the brevity of the brief, but the criteria applied to it and the division of labour around it.

Humans review the brief, and only the brief carries weight. Reviewers ask three questions: Is this a worthwhile pursuit? Is the approach convincing? Are the likely gains worth the effort? This is the blunt approach research funders have taken for decades, and these are precisely the judgment calls at which machines are worst and humans best. The appendix, by contrast, is verified by models rather than reviewed by humans. AI-written content is checked by AI that recomputes the statistics, re-executes the code and re-queries the data, and produces a checklist. That checklist is placed in front of the human reviewer alongside the brief.

None of this is exotic. Highly selective journals like \emph{Nature} and \emph{Science} already pair a short main text with a long supplement, and their editors judge importance from the short document. These venues curate openly: acceptance has never really meant ``this is correct'' so much as ``this matters.'' Accepting a paper becomes a curatorial act, while the archived object remains available in full for anyone who wants to build on it.

\section{Discussion}
The sharpest objection is that this rewards coolness: humans grade a pitch while the methodological detail is filed under ``supplementary'' and skipped. But judging whether a pursuit is worthwhile is itself a judgment of substance. What moves to the machines is the experimental record, and acceptance requires passing both gates. A seductive brief whose claims cannot be traced into a sound, verifiable appendix fails verification, and re-running an analysis is a far stronger check on rigour than a tired reviewer skimming a methods section on faith.

Nothing would stop a reviewer from feeding a brief into a model, of course. Most likely, reviewers already do this with twenty-page papers. Our proposal makes that workload manageable again, and a reviewer cast as curator has reason to answer a handful of questions about a brief themselves.

A further criticism is that ``worthwhile'' is more subjective than the criteria it replaces, and that a model could choose goals for the field as rationally as humans can. Both are true, and neither is a counter-argument. Review is already saturated with taste dressed up as soundness, and making the judgment of worth explicit makes it contestable. As for rationality, the choice of goals was never about rationality but about values, and values belong to the people who must live with the outcomes. That is not an authority we should want to hand over.

Nor does the proposal kill scholarship whose contribution is carefully crafted prose. An essay goes into the appendix whole, exactly as its author wrote it, and the brief states what it argues and why that matters.

Another criticism is that our proposal will accelerate the homogenisation of papers. We believe a prescribed 1500-word format standardises less, not more. Today each venue's house formula polices every page of a submission. Under our proposal the uniform surface shrinks to a short brief, and the appendix escapes style-grading altogether. And a brief is something practitioners and policymakers will actually read, and could realistically write.

\section{Limitations and Conclusion}
This paper is a provocation, and the proposal is deliberately under-specified. How many reviewers should judge a brief, how ``worthwhile'' is calibrated across reviewers and venues, who audits the verifiers: these are questions for testing and iteration.

In short, we need to separate the writing from the reviewing, let machines verify what machines can verify, and spend scarce human judgment on the one question that still demands it: where should this field go next? We propose papers made of a 1500-word brief curated by humans and an unlimited appendix verified by models.

Three disclosures, in the paper's own spirit. We cite no sources: every claim here is offered as an argument to be weighed on its face. Portions of this text were drafted with a language model, and under our own proposal you should not care: judge the pursuit, not the prose. And this document is itself under 1500 words.

\end{document}